\documentclass[twocolumn,preprintnumbers,amsmath,amssymb]{revtex4}
\usepackage{graphicx}
\usepackage{dcolumn}
\usepackage{bm}
\begin{document}
\title{
Electrically-induced n-i-p junctions  in 
multiple graphene layer structures
}
\author{
M.~Ryzhii\footnote{Electronic mail: m-ryzhii(at)u-aizu.ac.jp} and V.~Ryzhii
}
\affiliation{
Computational Nanoelectronics Laboratory, University of Aizu, 
Aizu-Wakamatsu 965-8580 and\\
Japan Science and Technology Agency, CREST, Tokyo 107-0075, Japan
}
\author{T. Otsuji}
\affiliation{
Research Institute of Electrical Communication, Tohoku University, Sendai 980-8577 and\\
Japan Science and Technology Agency, CREST, Tokyo 107-0075, Japan
}
\author{V.~Mitin}
\affiliation{
Department of Electrical Engineering,
University at Buffalo, State University of New York, NY 14260, USA
}
\author{M.~S.~Shur}
\affiliation{
Department of Electrical, Computer, and Systems Engineering,
Rensselaer Polytechnic Institute, Troy, New York 12180, USA
}
\date{\today}

\begin{abstract}
The Fermi energies of electrons and holes and their densities in different
graphene layers (GLs) in the n- and p-regions of the electrically induced
n-i-p junctions formed in  multiple-GL structures are calculated 
both numerically
and using a simplified analytical model.
The reverse current associated with the injection of minority
carriers through the n- and p-regions
in the electrically-induced  n-i-p junctions under the reverse bias
is calculated as well.
It is shown that  in the electrically-induced  n-i-p junctions with 
moderate  numbers of GLs 
the reverse current can be 
substantially suppressed. 
Hence,  multiple-GL structures with such n-i-p junctions can be used
in different electron and optoelectron devices.
\end{abstract}

\maketitle

\begin{figure}[h]\label{Fig.1}
\begin{center}
\includegraphics[width=6.9cm]{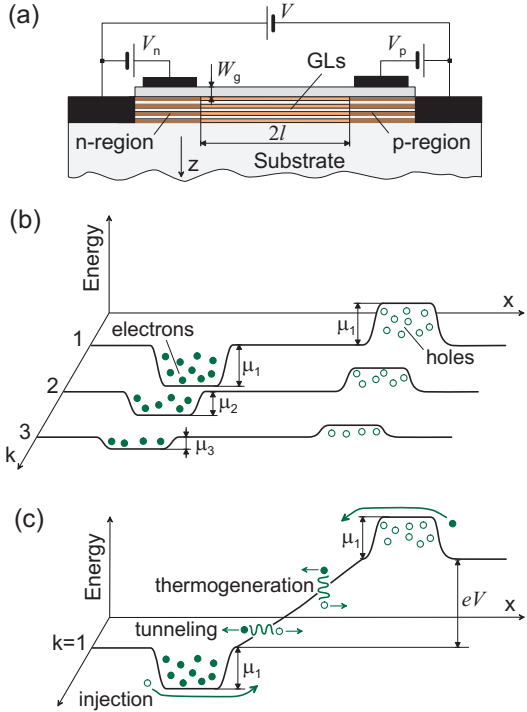}
\caption{Schematic view  of (a) a multiple-GL structure
and its band diagrams at (b) zero bias voltage (V = 0)
and (c) at reverse bias (for GL with $k = 1$). Arrows indicate
the directions of propagation of injected electrons and holes as well as
those thermogenerated and generated due to interband tunneling.
}
\end{center}
\end{figure}

\section{Introduction}
The possibility to form electrically-induced  n-p and n-i-p 
junctions~\cite{1,2,3}
in gated graphene layers (GLs), as well as lateral arrays of graphene 
nanoribbons and graphene bilayer opens up prospects of creation
novel electronic and optoelectronic devices
~\cite{4,5,6,7,8}. In contrast to GL structures with chemically
doped n- and p-region 
in the GL structures with  electrically-induced n-p and n-i-p junctions, 
there is a possibility of their voltage control.
 Recent success in fabricating high-quality multiple GLs~\cite{9,10}
stimulates an interest in different prospective 
devices in which the utilization
of  multiple-GL structures (stacks of disoriented GLs)
 instead of single-GL structures
can provide a significant improvement
of their performance. Such devices include, in particular,  
terahertz tunneling transit-time oscillators (similar to
 that considered in Ref.~\cite{4}),
lasers with optical
and electrical pumping, and  high performance interband
photodetectors~\cite{11,12,13}.
Gated multiple-GL structures can be also used in high-frequency
field-effect transistors~\cite{14} and other devices 
(the terahertz frequency multipliers, and plasmonic devices). 
However  the penetration  of the electric field (transverse to the GL plane) 
beyond the topmost GL
 as well as its sceening by electron or hole charges
in GLs can substantially limit the influence of the gates
(the effect of the quantum capacitance~\cite{15}).
In this paper, we study the influence of screening in
 gated multiple-GL structures on the formation and characteristics
 of n- or p-regions
and n-i-p junctions
in these structures. We calculate the electron and hole Fermi energies
and densities in GLs in the n- and p-regions as functions of the GL index,
gate voltage, and temperature. Using these data, we find the voltage and
temperature 
dependences of the reverse current in
the n-i-p junctions with different structural parameters.

\section{Equations of the model}

Let us consider a multiple-GL structure with the side Ohmic contacts
to all GLs
and two split gates (isolated from GLs)  on the top of this structure
as shown in Fig.~1(a).
Applying the 
positive ($V_n = V_g > 0$) or negative  ($V_p= - V_g < 0$) voltage
between the gate and the adjacent contact (gate voltage), one can obtain the electrically-induced n- or p-region. 
In the single- 
and multiple-GL structures with
 two split top
gates under the voltages of different polarity, one can create
lateral n-p or n-i-p junctions. 
Generally, the source-drain voltage $V$ can be applied between the side Ohmic
contacts to GLs. Depending on the polarity of this voltage,
the n-p and n-i-p junctions can be either  direct or reverse  biased. 
We  assume that the potentials of the first (source) contact and the pertinent
gate
are $\varphi_s = 0$ and $\varphi_g = V_g > 0$, respectively, and
the potentials of  another gate and contact (drain) are
$\varphi_g = - V_g < 0$  and $\varphi_d = V = 0$ (or  $\varphi_d = V \neq  0$.
If the slot between the gates $2Lg$ is sufficiently wide (markedly exceeds the thickness of the gate layer $W_g$
separating the gates and the topmost GL), there are intrinsic i-regions
in each GL under the slot.
Thus the n-i-p junction is formed.
The pertinent band diagrams are shown in Fig.~1(b) and 1(c).
Since the side contacts are  the Ohmic contacts, 
the electron Fermi energy in the $k$-th GL
sufficiently far from the contacts are given by $\mu_k = \pm e\varphi_k$.
Here $e$ is the electron charge and $\varphi_k = \varphi|_{z = kd}$ 
is the potential of the $k$-th GL, 
$k = 1,2,...,K$,
where $K$ is the number of GLs in the structure, $d$ 
is the spacing between GLs,  and the axis $z$ is directed perpendicular to the GL plane with $z = 0$ corresponding to the topmost GL and $z = z_K =
Kd$ - to the lowest one.

Focusing on the n-region (the p-region can be considered in a similar way)
and introducing the dimensionless potential 
$\psi = 2\varphi/V_g$, one can arrive at the following
 one-dimensional
Poisson equation governing the potential distribution in the $z$-direction
 (in the n-region): 
\begin{equation}\label{1}
\frac{d^2\psi}{dz^2} = \frac{8\pi\,e}{\ae\,V_g}\sum_{k=1}^K(\Sigma^{-}_k
- \Sigma^{+}_k)\cdot
\delta(z - kd +d).
\end{equation}
Here $\ae$ is the dielectric constant, $\delta (z)$ and $\Sigma^{-}_k$
and $\Sigma^{+}_k$
are the equilibrium sheet densities  
in the $k$-th GL of electrons and holes, respectively.
These densities, taking into account the linear dispersion low for electrons
and holes in graphene,  are expressed via the electron Fermi energy as

$$
\Sigma^{\mp}_k =  \frac{2}{\pi}\biggl(\frac{k_BT}{\hbar\,v_F}\biggr)^2\int_0^{\infty}\frac{d\xi\,\xi}{1 + \exp{(\xi \mp \mu_k/k_BT)}}
$$
\begin{equation}\label{2}
= \frac{12\Sigma_T}{\pi^2} \,\int_0^{\infty}\frac{d\xi\,\xi}{1 + 
\exp(\xi \mp \mu_k/k_BT)},
\end{equation}
where $\Sigma_T = (\pi/6)(k_BT/\hbar\,v_F)^2$ is the electron and
hole density in the intrinsic graphene at the temperature $T$,
$v_F \simeq 10^8$~cm/s is the characteristic velocity of electrons and holes
in graphene, and $\hbar$ and
$k_B$ are the Planck and Boltzmann constants, respectively.
Here it is assumed that the electron (hole) energy spectrum is
$\varepsilon = v_Fp$, where $p$ is the absolute value of the electron
momentum. 
The boundary conditions are assumed to be as follows:
\begin{equation}\label{3}
\psi|_{z = -0} = 2 + W_g\frac{d \psi}{dz} \biggr|_{z = -0},\qquad
\frac{d \psi}{dz} \biggr|_{z = kd + 0} = 0.
\end{equation}

\begin{figure}[t]
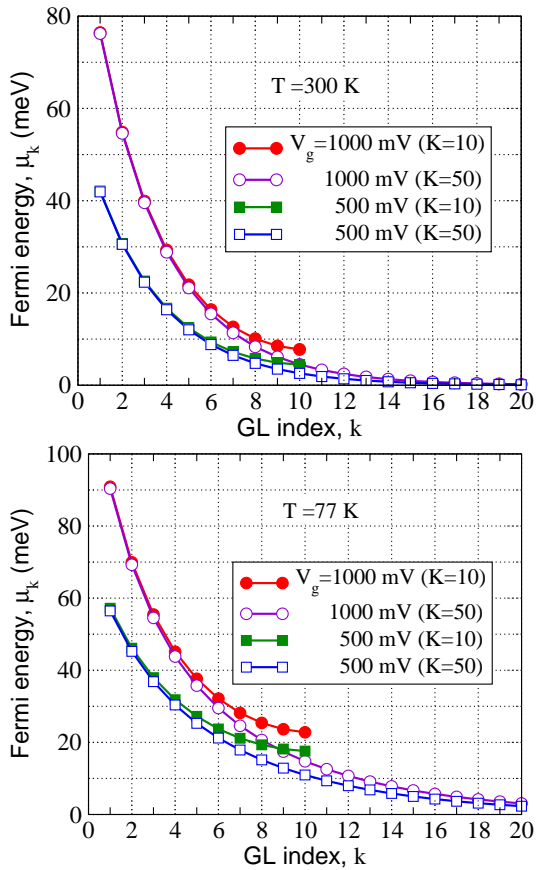
\label{Fig.2}
\begin{center}
\includegraphics[width=7.0cm]{SCREEN_F2a.eps}
\includegraphics[width=7.0cm]{SCREEN_F2b.eps}
\caption{ The electron Fermi energy $\mu_k$
as a function of the GL index $k$ calculated for 
multiple-GL structures with different number of GLs $K$
for different gate voltages $V_g$
at  $T = 300$~K (upper panel) and  $T = 77$~K (lower panel).
}
\end{center}
\end{figure}   

\begin{figure}[t]\label{Fig.3}
\begin{center}
\includegraphics[width=7.0cm]{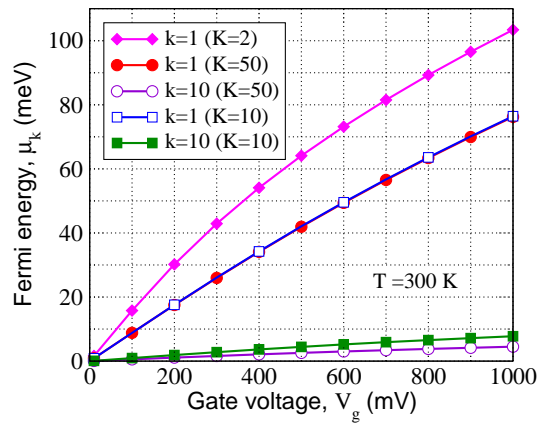}
\caption{ Voltage dependences of  electron Fermi energies 
in some GLs 
in   multiple-GL structure with different $K$ 
at  $T = 300$~K. 
}
\end{center}
\end{figure}


Equations~(1) - (3)  yield
\begin{equation}\label{4}
2 - \psi_1 = \Gamma\Phi(\psi_1),
\end{equation}
for $K = 1$,

$$
\frac{d}{W_g} (2 - \psi_1) - \psi_1 + \psi_2 = \frac{d}{W_g}\Gamma \Phi(\psi_1),
$$
\begin{equation}\label{5}
\psi_{1} - \psi_2 = \Gamma \Phi(\psi_2)
\end{equation}
for $K = 2$,
and
$$
\frac{d}{W_g} (2 - \psi_1) - \psi_1 + \psi_2 = \frac{d}{W_g}\Gamma \Phi(\psi_1),
$$
$$
\psi_{k-1}  - 2\psi_{k} + \psi_{k+1} = \frac{d}{W_g}\Gamma \Phi(\psi_k), \qquad  
(2 \leq  k \leq  K-1),
$$
\begin{equation}\label{6}
\psi_{K-1} - \psi_K = \frac{d}{W_g}\Gamma \Phi(\psi_K)
\end{equation}
for $K > 2$.
Here 
$$
\Phi(\psi) = \frac{12}{\pi^2}\biggl[
\int_0^{\infty}\frac{d\xi\,\xi}{1 + \exp(\xi - U_g\psi)}
$$
\begin{equation}\label{7}
- \int_0^{\infty}\frac{d\xi\,\xi}{1 + \exp(\xi + U_g\psi)}\biggr],
\end{equation}
where
$\Gamma = (8\pi/\ae)(eW_g\Sigma_T/V_g) \propto T^2/V_g$ 
and $U_g = eV_g/2k_BT$.

\section{Numerical results}

Equations~(4) - (7) were solved numerically.
The results of the calculations are shown in Figs.~2 - 5.
In these calculations,
we assumed that $\ae = 4$, $d = 0.35$~nm, and $W_g = 10$~nm.

Figure~2 shows the dependences of the electron Fermi energy 
\begin{equation}\label{8}
\mu_k   = \frac{eV_g}{2}\,\psi_k
\end{equation}
as a function of the GL index $k$ calculated for 
multiple-GL structures with different number of GLs $K$
at  different gate voltages and temperatures.
One can see that the Fermi energy steeply decreases with increasing GL index.
However, in GLs with not too large $k$, the Fermi energy is larger or about
of the thermal energy. 
As one might expect, the electron Fermi energies in all GLs at $T = 77$~K
are somewhat larger than at $T = 300$~K (see also Fig.~5).
The obtained values of the electron Fermi energies in topmost GLs 
are $\mu_1 \simeq 92$~meV  and $\mu_1 \simeq 77$~meV for $V_g = 1000$~mV
at  $T = 77$~K
and  $T = 300$~K, respectively.

Figure~3 shows the voltage dependences of the electron Fermi energies
in some GLs
in  multiple-GL structure with $K = 2$, $K = 10$, and $K = 50$
at $T = 300$~K. 

Figure~4 shows the electron densities $\Sigma_k^{-}$ in the structures with
different number of GLs $K$ at different temperatures.
One can see that the calculated 
electron densities in GLs with sufficiently
large indices ($k > 15$ at   $T = 77$~K and $T = 300$~K)
approach  to their values in the intrinsic graphene 
($\Sigma_T = 0.59\times 10^{10}$~cm$^{-2}$
 and $8.97\times 10^{10}$~cm$^{-2}$).
The electron densities in GLs in the structures with different $K$
are rather close to each other, particularly, in GLs with small and moderate
indices.

Figure~5 presents the Fermi energies in GLs with different indices
at different temperatures. One can see from Fig.~5 (as well as from Fig.~2)
that the higher $T$
corresponds to lower  $\mu_k$. This is due to an increasing dependence
of the density of states on the energy and the thermal spread in the electron energies.

\section{Analytical model} 

\begin{figure}[t]\label{Fig.3}
\begin{center}
\includegraphics[width=7.3cm]{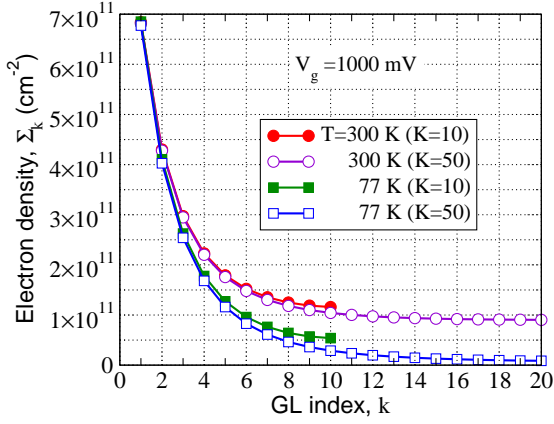}
\caption{ Electron density vs   GL index in multiple GL-structures with
different number of GLs ($K = 10$ and $K = 50$) at different temperatures
and $V_g = 1000$~mV. 
}
\end{center}
\end{figure}  

\begin{figure}[t]\label{Fig.5}
\begin{center}
\includegraphics[width=7.0cm]{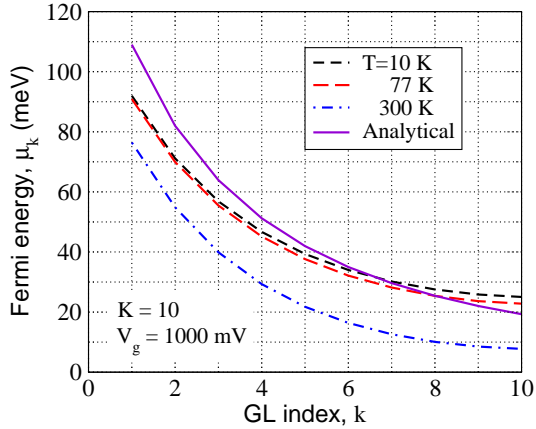}
\caption{Comparison of the $\mu_k$ vs $k$ dependences calculated
for different temperatures
using numerical and  simplified analytical (solid line)
models. 
}
\end{center}
\end{figure}
At not too low gate voltages when $U_g \gg 1$, one can assume that in
 a number of GLs
 the electrons  under the gate are degenerate,
i.e., $\mu_k \gg k_BT$, 
and the contribution of holes (nondegenerate) can be disregarded,
hence,  from Eq.~(7) we obtain

\begin{equation}\label{9}
\Phi(\psi) \simeq \frac{6}{\pi^2}U_g^2\psi^2.
\end{equation}
In this case, for a single-GL structure ($K = 1$), Eq.~(4)
yields
\begin{equation}\label{10}
2 - \psi_1 \simeq \frac{6}{\pi^2}\Gamma\,U_g^2\psi_1^2,
\end{equation}
Solving Eq.~(10) and considering Eq.~(8), for the electron Fermi
energy in a GL (in a single-GL structure),
we obtain

\begin{equation}\label{11}
\mu_1 \simeq \mu_g\sqrt{1 - \frac{\mu_g}{eV_g}},
\end{equation}
where
$\mu_g = \hbar\,v_F\sqrt{\ae\, V_g/4eW_g}$.
For the same parameters as those used in Figs.~2- 4,
$\mu_1 \simeq 150$~meV.
Such a  value is  markedly  larger than those  
calculated for the topmost GLs in
multiple-GL structures (although it is somewhat exaggerated because
the temperature spread in the electron energies is disregarded). 
This can be attributed to the fact
that in multiple-GL structures the  electron density is shared between
the topmost GL and underlying GLs resulting in lower Fermi energies in
all of them.

Considering  multiple-GL structure with large $K$, one can
neglect the discreteness of the structure
and replace the summation in  Eq.~(1) by the integration. As a result,
following Ref.~\cite{13},
 one can arrive at
\begin{equation}\label{12}
\frac{d^2\psi}{dz^2} = 0 \qquad (-W_g < z < 0,\, z > z_K),
\end{equation}
\begin{equation}\label{13}
\frac{d^2\psi}{dz^2} = \frac{\Gamma}{dW_g}\Phi(\psi)\qquad (0 < z < z_K).
\end{equation}

In this case, 
considering Eq.~(7),
we arrive at 
\begin{equation}\label{14}
\frac{d^2\psi}{dz^2} = \frac{\psi^2}{L_s^2}
\end{equation}
with the characteristic screening length
\begin{equation}\label{15}
L_s = \frac{\pi\,\sqrt{dW_g}}{\sqrt{6\Gamma}U_g} =
\hbar\,v_F\sqrt{\frac{\ae\,d}{2e^3V_g}} \propto V_g^{-1/2}.
\end{equation}
The boundary conditions for Eq.~(14) are
\begin{equation}\label{16}
\psi|_{z = 0} = 2 + W_g\frac{d\psi}{dz}\biggr|_{z = -0}, 
\qquad \frac{d\psi}{dz}\biggr|_{z =  z_K+0} = 0,
\end{equation}

In multiple-GL structures with a large number of GLs ($K \gg 1$),
one can extend the coordinate of the lowest GL to infinity and set 
$d\psi/dz|_{z =\infty} = 0$ with
 $\psi|_{\infty} = 0$.
Solving Eq.~(14) with the latter boundary conditions,
we arrive at
\begin{equation}\label{17}
\psi = \frac{1}{(C + z/\sqrt{6}L_s)^2},
\end{equation}
where $C$ satisfies the following equation:
\begin{equation}\label{18}
C^3 - C/2 = (W_g/\sqrt{6}L_s), 
\end{equation}
%

Since in reality $W_g \gg \sqrt{6}L_s$, one obtains 
$C \simeq (W_g/\sqrt{6}L_s)^{1/3} \propto V_g^{1/6}$


Taking into account Eq.~(8),
Eq.~(17) yields
\begin{equation}\label{19}
\mu_k \simeq  \frac{eV_g}{2[C + (k - 1)d/\sqrt{6}L_s]^2}
= \mu_1\,a_k.
\end{equation}
Here 
\begin{equation}\label{20}
\mu_1  = \frac{eV_g}{2C^2} 
\propto \biggl(\frac{V_g}{W_g}\biggr)^{2/3}
\end{equation}
is the Fermi energy of electrons
in the topmost GL in the n-section  (holes in the p-section),
\begin{equation}\label{21}
a_k = [1 + (k - 1)\gamma]^{-2},
\end{equation}
and 
\begin{equation}\label{22}
\gamma = \frac{d}{\sqrt{6}L_sC} \propto 
\frac{d}{W_g^{1/3}L_s^{2/3}}
\propto \biggl(\frac{V_g}{W_g}\biggr)^{1/3}.
\end{equation}
%
Setting $d = 0.35$~nm, $W_g = 10$~nm, $\ae = 4$,  and $V_g = 1$~V, one can
obtain $L_s \simeq 0.44$~nm. 
$C \simeq 2.14$, $\mu_1 \simeq 109~meV$, and $\gamma \simeq 0.153$.
The $\mu_k$ versus $k$ dependence obtained using our simplified
analytical model, i.e.,  given by Eqs.~(15) and (17)
is shown by a solid line in Fig.~5.

Equations~(9) - (22) are valid when $\mu_K > k_BT$, i.e., at sufficiently
large $V_g$ or/and sufficiently small $T$. 
Since the Fermi energy (at fixed electron density) decreases with increasing
temperature, the above formulas of our simplified (idealistic)
 model yield somewhat
exaggerated values of this energy [compare the dependences in Fig.~5 obtained
numerically using Eqs.~(4) - (7) and that found analytically using 
Eqs.~(19)- (22)].

\section{The reverse current}

The current across the n-i-p junctions under their reverse bias
($V < 0$, see Fig.~2(c))
is an important characteristic of such junctions~\cite{16}.
In particular, this current
can substantially affect the performance of 
the terahertz tunneling transit-time oscillators 
and interband photodetectors~\cite{5,13}.
This current is associated with the thermogeneration and tunneling
generation of the electron-hole pairs in the i-region.
A significant contribution to this current can be provided by
the injection of minority carriers (holes in the n-region and electrons in the p-region). Such an injection current in the $k$-th GL is determined
by the height of the barrier for the minority carrier which, in turn,  is
determined by the Fermi energy $\mu_k$ of the majority carrier.
The latter, as shown above,  depends on the gate voltage and the GL index.
As a result, the reverse current can be presented as
\begin{equation}\label{23}
J = J_i + J_{th} + J_{tunn},
\end{equation}
where the injection current (which is assumed to be of the thermionic origin)
 is given by
$$
J_i = 
\frac{2ev_F}{\pi^2}\biggl(\frac{k_BT}{\hbar\,v_F}\biggr)^2
\int_{-\pi/2}^{\pi/2}d\theta\cos\theta
$$
$$
\times\sum_{k = 1}^{K}
\int_{0}^{\infty}
\frac{d\xi\xi}{1 + \exp(\xi + \mu_k/k_BT)} 
$$
$$
= \frac{24ev_F\Sigma_T}{\pi^3}
\sum_{k = 1}^{K}
\int_{0}^{\infty}
\frac{d\xi\xi}{1 + \exp(\xi + \mu_k/k_BT)}. 
$$
\begin{equation}\label{24}
= \frac{12J_T}{\pi^2}
\sum_{k = 1}^{K}
\int_{0}^{\infty}
\frac{d\xi\xi}{1 + \exp(\xi + \mu_k/k_BT)}
\end{equation}
with $J_T = 2ev_F\Sigma_T/\pi$. At $T = 77 - 300$~K, $J_T 
\simeq 0.06 - 0.9$~A/cm.
Deriving Eq.~(24), we have taken into account that the distribution function
of holes  which enter the
 n-region overcoming the barrier in the $k$-th GL with the height $\mu_k$
 is $f_k^{pn} \simeq \{1 +  \exp[(\mu_k +v_Fp)/k_BT]\}^{-1}$. 
Similar formula 
is valid for electrons in the p-region. The factor $2$ appears in Eq.~(24)
due to the contribution of both holes and electrons.
Equation~(24) is valid when the bias voltage is not too small: $eV > k_BT$.
At $V \rightarrow 0$ one has $J_i \rightarrow 0$ 
(as well as $J_{th}$ and $J_{tunn}$).
Scattering of holes in the n-region and electrons in the p-region resulting
in returning of portions of them back to the contacts leads to some
 decrease in $J_i$.

The temperature dependences of the
reverse current associated with the injection
from the n- and p-region
calculated using Eq.~(24) (with
 the quantities
$\mu_k$   shown in  Fig.~5)
are presented in Fig.~6.
As one might expect, the reverse current sharply 
increases with the temperature and the number of GL.
The latter is due to relatively low energy barriers for minority
carriers in the n- and p-regions in GLs with large indices.
For comparison, the injection current in a single-GL structure
calculated using Eq.~(24) with Eq.~(8), is given by 
$$
J_i^{S} = \frac{12J_T}{\pi^2}
\int_{0}^{\infty}
\frac{d\xi\xi}{1 + \exp(\xi + \mu_1/k_BT)}
$$
\begin{equation}\label{25}
\simeq  \frac{12J_T}{\pi^2}
\exp\biggl(- \frac{\hbar\,v_F\sqrt{ \ae\,V_g/8eW_g}}{k_BT}\biggr).
\end{equation}
At the same parameters as above and  $T = 300$~K,  Eq.~(25) 
 yields $J_i^{S} \simeq 0.01$~A/cm.

\begin{figure}[t]\label{Fig.6}
\vspace*{-0.4cm}
\begin{center}
\includegraphics[width=7.0cm]{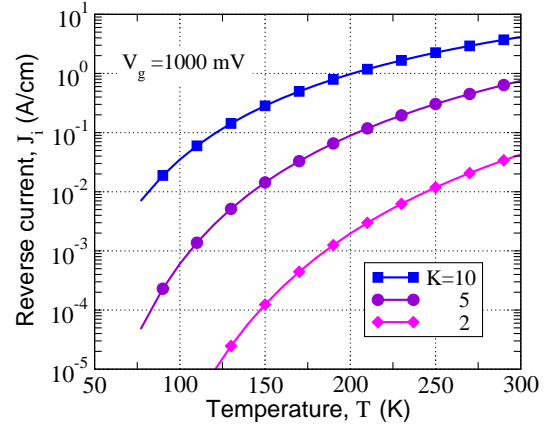}
\caption{Temperature dependence of reverse current  (injection component, $J_i$)
for multiple-GL structures with different number of GLs.
}
\end{center}
\end{figure}

%
%
%

When  $K \gg \gamma^{-1}\sqrt{\mu_1/k_BT}$,
the main contributions to the reverse current is associated with
GLs with large indices (in which the barriers are very low), so that
one obtains
\begin{equation}\label{26}
J_i \lesssim KJ_T.
\end{equation}
As follows from Eq.~(26), the injection current (and, therefore, the net
reverse current) can be fairly large due to the ``shortcut'' by GLs with
large indices (placed deep below the gate).

Since the thermogeneration is associated primarily with the absorption of optical phonons~\cite{17}, the pertinent rate $g_{th}$
is independent of the electric field
in the i-region, but it is proportional to the i-section length $2l$
($l \lesssim L_g$). 
The contribution of the thermogeneration to the reverse current can be presented as
\begin{equation}\label{27}
J_{th} = 4Kelg_{th}.
\end{equation}
The quantity $g_{th}$ as a function of the temperature was calculated
in Ref.~\cite{17}.
Equations~(24) - (27) are valid if $2l < l_R$, 
where $l_R$ is the recombination
length. In the situations when the bias voltage between the side contacts
is not too small (as it should be, for instance, in GL-based interband photodetectors), 
the recombination length is fairly long. Indeed, assuming
that the recombination time (at $T = 300$~K)
and the drift velocity 
  are
 $\tau_R = 5\times10^{-10}$~s~\cite{17} and 
$<v> \sim  v_F/2 = 5\times10^7$~cm/s~\cite{18}, respectively,
one obtains $l_R \simeq 250~\mu$m.

Assuming that  $2l = 10~\mu$m with $g_{th} = 10^{13}$~cm$^{-2}$s$^{-1}$
and  $g_{th} = 10^{21}$~cm$^{-2}$s$^{-1}$ at $T = 77$~K and $T = 300$~K,
respectively~\cite{17}, we obtain $J_{th} \simeq 3.2\times(10^{-9} -
10^{-1})$~A/cm. 
One can see that the thermogeneration contribution
to the reverse current is much smaller than the injection
contribution at lower temperatures, while it can be substantial
at $T = 300$~K in the n-i-p structures with long i-region.

The tunneling generation can  significantly contribute to the reverse
current
at elevated  electric fields in the i-region, i.e., in relatively short
GL structures at elevated
bias voltages~\cite{1,5}.
This current can be calculated using the following formula
which follows from the expression for the tunneling probability in 
GLs~\cite{1,2}
(see, for instance~\cite{5}):
\begin{equation}\label{28}
J_{tunn} = \frac{ev_F}{\pi^2l^2}\biggl(\frac{eV}{2l\hbar\,v_F}\biggr)^{3/2}
\propto \frac{V^{3/2}}{l^{1/2}}.
\end{equation}
Depending on the n-i-p junction applications,
the quantities $V$
and $l$ should be chosen to provide either domination of
 tunneling current 
(as in tunneling transit-time oscillators~\cite{4}) or its suppression
(as in the interband photodetectors~\cite{13}).


%

%


\section{Conclusions}
We calculated the dependences of 
the Fermi energies and densities
of electrons and holes 
 in the n- and p-regions of the electrically induced
n-i-p junctions formed in  multiple-GL structures 
on the GL indices,  gate voltage, temperature, and the structural parameters.
Using the obtained values of the Fermi energies and, hence, of the heights
of potential barriers for minority carriers  in 
 the n- and p-regions, we found the temperature dependences
of the reverse injection current for multiple-GL structures with different
numbers of GLs. It was shown that the formation
of  effective electrically-induced  n- and p-regions and n-i-p junctions, i.e.,
the n-i-p junctions with suppressed reverse currents
 in multiple-GL structures with several GLs is possible. 
The   utilization of the  electrically-induced   n-i-p junctions 
in multiple-GL structures 
in different devices, such as  
terahertz tunneling transit-time oscillators,
lasers, high performance interband
photodetectors, and some others
 might provide an enhancement of the device performance 
(an increase in output power and responsivity)
and widening of their functionality 
(owing to the possibility of the gate-voltage control).

\section*{Acknowledgments}
This work was supported 
by  the Japan Society for Promotion of Science
and by the Japan Science and Technology Agency, CREST, Japan.

\end{document}